\documentclass[twocolumn,prx,aps,superscriptaddress]{revtex4}
\usepackage{amsmath,amssymb}
\usepackage[colorlinks, citecolor=blue]{hyperref}
\usepackage{amsfonts}
\usepackage{mathrsfs}
\usepackage{graphicx}
\usepackage{amssymb}
\usepackage{color}
\usepackage{dcolumn}
\usepackage{bm}
\usepackage{tikz,pstricks}
\usepackage[header,title,page,titletoc]{appendix}
\usepackage{url}

\pgfarrowsdeclaredouble{feather}{feather}{to reversed}{to reversed}

\begin{document}

\title{Critical behavior of QED$_3$--Gross-Neveu-Yukawa Theory in an Arbitrary Gauge}
\author{Jiang Zhou}
\email{jzhou5@gzu.edu.cn}
\affiliation{Department of Physics, Guizhou University, Guiyang 550025, PR China}
\author{Su-Peng Kou}
%\thanks{Corresponding author}
%\email{spkou@bnu.edu.cn}
\affiliation{Center for Advanced Quantum Studies, Department of Physics, Beijing Normal
University, Beijing 100875, China}

\begin{abstract}
The chiral QED$_3$--Gross-Neveu-Yukawa (QED$_3$-GNY) theory is a $2+1$-dimensional U(1) gauge theory with $N_f$ flavors of four-component Dirac fermions coupled to a scalar field.
For $N_f=1$, the specific chiral Ising QED$_3$-GNY model has recently been conjectured to be dual to
the deconfined quantum critical point that describes Neel--valence-bond-solid transition of frustrated quantum magnets on square lattice.
We study the universal critical behaviors of the chiral QED$_3$-GNY model in $d=4-\epsilon$ dimensions for an arbitrary $N_f$ . 
We calculate the boson anomalous dimensions, inverse correlation length exponent, as well as 
the scaling dimensions of nonsinglet fermion bilinear in the chiral QED$_3$-GNY model. 
The Pad$\acute{e}$ estimates for the exponents are obtained in the chiral Ising-, XY- and Heisenberg-QED$_3$-GNY universality class respectively.
We also establish the general condition of the supersymmetric criticality for the ungauged QED$_3$-GNY model. 
For the conjectured duality between chiral QED$_3$-GNY critical point and deconfined quantum critical point, 
we find the inverse correlation length exponent has a lower boundary $\nu^{-1}>0.75$, beyond which the Ising-QED$_3$-GNY--$\mathbb{C}$P$^1$ duality may hold.
\end{abstract}

\maketitle

\section{Introduction}
The theoretical study of critical behavior undergoes several main stages. The first stage refers to the
classical phase transition at finite temperature, the many-body systems exhibit continuous phase transition at some critical point upon lowing the temperature. Each phases of matter separated by the critical point can be characterized in terms of a local order-parameter which allows to classify and distinguish in different phases. For classical criticality, the dominant fluctuations degrees of freedom at the critical point only relevant to space fluctuations of order-parameter.
The second stage refers to the quantum phase transition at zero temperature, the ground
state of strongly correlated systems changes qualitatively at the quantum critical point (QCP)\cite{qpt1}, upon tuning nontemperature parameters such as coupling strength and chemical doping. Just as the classical criticality, the quantum phase transition  have a classical analog and can be also characterized in terms of a local order-parameter. Unlike the case in classical criticality, the dominant fluctuations at QCP include both space and time fluctuations of bosonic order-parameter. Based on the Landau-Ginzburg symmetry-breaking theory and the concepts of scaling laws and universality, the renormalization group (RG) theory provides a satisfactory description for critical phenomena\cite{rg1,rg2,rg3}--this refers to as Landau-Ginzburg-Wilson (LGW) paradigm.

In recently years, the study of critical behaviors in two spatial dimensions are going through the exciting third stage at which the LGW approach fails. A paradigmatic example of such a phase transition is the Neel--valence-bond-solid (VBS) transition for Heisenberg spin-1/2 quantum antiferromagnets on the square lattice\cite{dqcp1,dqcp2}. The Neel phase corresponds to conventional antiferromagnetic phase which breaks spin-rotation symmetry spontaneously, the VBS phase corresponds to a conventional crystal of spin singlets which breaks lattice-rotational symmetry spontaneously. By fine tuning, it has been argued that the Neel-VBS transition is continuous and the QCP is governed by fractionalized bosonic spinon degrees of freedom coupled to a noncompact U(1) gauge field\cite{dqcp1,dqcp2}.
Both the fractionalized spinons and the gauge photon are confined in either phase, that is, they emerge only at the QCP. Hence, the phase transition has been dubbed deconfined QCP\cite{dqcp2,decon1}, and the corresponding dual description is given by the (2+1)-dimensional (2+1D) noncompact $\mathbb{C}$P$^1$ (NC$\mathbb{C}$P$^1$) model. Numerical studies on the on quantum magnets shows a striking continuous phase transition\cite{cptn1,cptn2,cptn3,cptn4}, which agrees with the theoretic expectations.

Such LGW-forbidden QCP can, however, occur between unconventional long-range entangled states and the conventional ordered states, i.e., transition from spin liquid to VBS order\cite{und}, and more examples have been reviewed by Ref.~[\onlinecite{und}]. More recently, various similar QCP have been suggested, i.e., transition between easy-plane XY antiferromagnets and VBS order\cite{decon1,numed}, between phases with competing order\cite{com1,com2,com3}, between two length scales\cite{twol}.

Apart from the deconfined QCP, there exist another category of QCP which includes the gapless critical fermionic degrees of freedom coupled to bosonic fluctuations via Yukawa-type interaction. Owing to gapless fermionic fluctuations, the theoretical description of QCP is given by the chiral Gross-Neveu-Yukawa (GNY) model\cite{gnydop,gnyne,gnythl,gnyfi,fourf,gnyfol}, or its purely fermionic version, the Gross-Neveu model\cite{gnythl}.
To understand the critical properties of such QCP better in three dimensions, the chiral Gross-Neveu model has been studied by many field-theoretical methods, including perturbative RG\cite{perrg,rosen}, large $N$ expansion\cite{largn,largn1}. The critical behavior of the chiral GNY model have recently been evaluated at three-loop order\cite{gnythl}, four-loop orders\cite{gnyfol,fourlgn} and large $N$ formalism\cite{largnh}.
More, a comprehensive analysis was carried out in [\onlinecite{compreana}].
This category of QCP always occurs in the interacting systems with Dirac and Weyl excitations\cite{qchub,qcgra,unive,mottd}, such as graphene superconducting criticality\cite{sccri},  emergent supersymmetric criticality on the surface of three-dimensional (3D) topological insulators\cite{stsup,stsup1}, fermion-induced quantum criticality in Weyl or Dirac semimetals\cite{fiqcp1,fiqcp2}.
It's believe that these QCP belong to different chiral GNY universality class.

The field-theoretical descriptions of the emergent properties at QCP is challenging. The NC$\mathbb{C}$P$^1$ gauge field theories give a natural route to a potentially continuous transition between two distinct symmetry-broken phases\cite{dqcp1,und}. It's argued that the Neel-VBS deconfined QCP differs from the O(3) Wilson-Fisher fixed point, and the existence of monopole event implies the breakdown of LGW paradigm. Alternatively, there exist another equivalent description of deconfined QCP directly in terms of the order-parameter by using nonlinear sigma model with a topological term\cite{sigma1,sigma2,sigma3}. For SU(2) Neel-VBS deconfined QCP on square lattice, the order-parameter is given by a Neel-VBS five-tuplet vector, and the topological term is given by Wess-Zumino-Witten term which directly leads to the breakdown of LGW paradigm at a deconfined QCP\cite{sigma3}. The sigma model description gave rise to the possibility that the QCP may have an enlarged symmetry, which rotates components of the five-tuplet vector into each other.

There exist, however, a proposed duality web between different descriptions of QCP\cite{dw1,dw2}, which provides an alternative perspective to understand QCP. The duality web builds on some earlier  fascinating work in the condensed matter and high energy physics\cite{dw1} and the equivalence between easy-plane NCCP$^1$ critical field theory in quantum magnets and 2+1D quantum electrodynamics (QED$_3$)\cite{ano1,ano2}. Building on these boson-fermion dualities, the duality web conjectured that the bosonic and fermionic description for deconfined QCP are dual to each other\cite{dw1}, all these boson or fermion fields are coupled to a dynamical U(1) gauge field. Via duality, two seemingly different theory may be infrared equivalent, namely, they describe the same long-distance behavior. The recent Monte Carlo study of the duality between easy-plane NC$\mathbb{C}$P$^1$ model and QED$_3$ theory with two flavors of two-component Dirac fermions has provided strong support for the duality conjecture\cite{numed}.

Given this recent development, another underlying description for exotic criticality is the chiral QED$_3$-GNY universality class. For Neel-VBS deconfined QCP, the NC$\mathbb{C}$P$^1$ critical field theory has been conjectured to be dual to chiral Ising QED$_3$-GNY model\cite{dw1}. An immediate consequence of the duality is the emergence of enlarged SO(5) symmetry for both theory at the deconfined critical point\cite{dw1}, which is supported by recent numerical evidences\cite{nume1,nume2}. The SO(5) symmetry also implies the coincidence of scaling dimensions between each component of five-tuplet. Besides, the existence of SO(5) symmetry in favor of self-duality of NCCP$^1$ theory.

From the purely theoretical aspects, the chiral QED$_3$-GNY theory bridges between condensed matter and particle physics.
Besides, it connects closely with some rich deconfined QCPs\cite{edgny2,edgny3,edgny4}. Examples of the chiral QED$_3$-GNY theory include criticality between algebraic and $Z_2$ spin liquid\cite{algz2},
algebraic and VBS transition\cite{algvbs}. The chiral GNY universality class has been extensively studied.
However, we have rare knowledge of the chiral QED$_3$-GNY critical behavior and absence of comprehensive understand of the properties of chiral QED$_3$-GNY universality class.
That is one aim of this paper to compute the critical exponent of QED$_3$-GNY theory.

The duality conjecture in (2+1)D, although having passed a number of consistency checks based on the symmetries\cite{dw1} and numerical evidence, still lacks of concrete proof. Various theoretical studies of chiral QED$_3$-GNY model have been carried out recently to test the prediction of duality conjecture. These include one-loop Wilson RG\cite{edgny2}, $\epsilon$ expansion up to three and four loops order\cite{edgny3,edgny4} and the $1/N$ expansion in fixed space-time dimension $d=3$\cite{nexpan}. Some of the critical exponents estimated from these studies (even at higher-loop orders), unfortunately, are incompatible with the numerical ranges and seem to be not inconsistent with the duality prediction. Therefore, the duality conjecture needs to be further confirmed in a reasonable way.

Motivated by these latest progresses, in this paper, we study the critical behavior of general chiral QED$_3$-GNY model for general number of fermion flavors $N_f$ in an arbitral gauge. We calculate the anomalous dimension and inverse correlation length exponent $\nu^{-1}$ of the QED$_3$-GNY theory in chiral Ising-, XY- and Heisenberg-universality class, respectively. The scaling dimensions of flavor-symmetry breaking nonsinglet fermion bilinear have also been evaluated to check the Ising-QED$_3$-GNY--CP$^1$ duality.

The paper is organized as follows. In the following section, we introduce the chiral QED$_3$-GNY theory under investigation. Sec.\ref{section3} specify the RG procedure and the employed techniques, the RG flow equations and the anomalous dimensions are also presented in this section. Sec.\ref{section4} is devoted to the computations of results for the critical exponents at leading order, Pad$\acute{e}$ approximants are then used for universal exponents in three dimensions spacetime. In Sec. \ref{section5}, we determine the scaling dimensions of adjoint nonsinglet fermion bilinear and then discuss the duality predictions.  Finally, brief conclusion are provided in Sec.\ref{section6}. More details about the calculations are presented in Appendix.

\section{model descriptions}\label{section2}
\subsection{The general chiral QED$_{3}$-GNY model}
The chiral QED$_{3}$-GNY model describes a variety of QCP in Dirac systems\cite{dw1,edgny2,edgny3,edgny4}. In this model, the Dirac fermionic fields
couple to the order parameters via Yuakawa coupling and the order parameters are described by a bosonic scalar fields. Here, we are interested in the general chiral QED$_{3}$-GNY
model, with the Lagrangian is defined by
$\mathcal{L}=\mathcal{L}_{QED_{3}}+\mathcal{L}_{\phi }+\mathcal{L}_{\psi \phi }$,
\begin{eqnarray}
\mathcal{L}_{QED_{3}} &=&\bar{\psi}_{i}(i\partial \!\!\!/ _{\mu }-eA \!\!\!/ _{\mu })\psi _{i}-\frac{1}{4}F_{\mu \nu }F^{\mu \nu },\label{e1} \\
\mathcal{L}_{\phi } &=&\frac{1}{2}(\partial _{\mu }\phi _{a})^{2}-\frac{m^2}{2}\phi _{a}\phi _{a}-\frac{\lambda }{4!}\left(\phi_{a}\phi _{a}\right)^{2}, \label{e2} \\
\mathcal{L}_{\psi \phi } &=&-g\bar{\psi}_{i}\left(\Sigma^{a}\phi _{a}\right) \psi _{i},\label{e3}
\end{eqnarray}%
in (2+1)D Minkowski time-space, where the notation $\partial \!\!\!/ =\gamma ^{\mu }\partial _{\mu }$ is Feynman slash. The repeated indices are summed over.
$\psi _{i}$ and its conjugate $\bar{\psi}_{i}$ are $2N_f$ flavors of two-component spinors, $i=1,2,...2N_f$, which can be combined into $N_f$ flavors of four-component spinors\cite{edgny4,nexpan}.
For generality, we will work with an arbitrary flavors number $N_{f}$ of four-component spinor in the following, then the $4\times 4$ gamma matrices $\gamma ^{\mu }$ form four-dimensional representation of the
Clifford algebra: $\{\gamma ^{\mu },\gamma ^{\nu }\}=2\delta^{\mu \nu }1_4$, with the indices $\mu $, $\nu =0,1,2$. The conjugate spinor is defined as $\bar{\psi}_{i}={\psi}_{i}^{\dag}\gamma_0$. The RG calculations below does not need the explicit matrix form of these gamma matrices. Indeed, we just need the dimensions and the commutativity or anticommutativity between them. The field strength tensor is $F_{\mu \nu }=\partial _{\mu }A_{\nu }-\partial_{\nu }A_{\mu }$, the gauge field coupled to fermionic fields through the minimal coupling. To check the gauge invariance, the gauge fixed-term
\begin{equation}
 \mathcal{L}_{\xi}= -\frac{1}{2\xi }(\partial_{\mu}A_{\mu})^{2},
\end{equation}
should be added in the Lagrangian. In the follows, we will work in an arbitrary gauge.

The Lagrangian given by $\mathcal{L}_{\phi }$ describes the pure bosonic fields, it includes kinetic part and quartic interactions with strength $\lambda $. To describe different universality class, we generalize the model by introducing a $N_{b}$-component vector for bosonic fields $\phi_a$, $a=1,2,...N_b$. The boson mass-square $m^{2}$
plays the role of tuning parameter for spontaneous symmetry breaking. In the symmetric phase ($m^{2}>0$), $\langle \phi _{a}\rangle =0$. In the symmetric
broken phase($m^{2}<0$), the scale field $\phi _{a}$ acquires a nonzero vacuum expectation, implying the spontaneous breaking of time-reversal symmetry and the dynamical
generation of a fermion mass.

$\mathcal{L}_{\psi\phi }$ denotes Yukawa coupling between bosonic and fermionic fields. The matrices $(\Sigma^{a})_{4N_f}$, signalled various breaking patterns of the gapped phases, entering in the fermion kinetic term are made mutually commuting, $[\Sigma^{a},\gamma ^{\mu}]=0$. In fact, the commuting relation always holds in the Mott criticality if $[\gamma^0,M^a]=0$, where $M^a$ is the mass matrix corresponding to the ordered phase with $\Sigma^a=\gamma^0 M^a$ and $\gamma^0$ is some suitable Dirac matrix\cite{com2,massm1,massm2}.
In this general QED$_{3}$-GNY model, the Yukawa couping $g$, the charge $e$ as well as the boson quartic interaction $\lambda$, are all marginal at upper critical dimension $d_{uc}=4$,
\begin{equation}
\left[ e^{0}\right] =\frac{4-d}{2},\left[ \lambda ^{0}\right] =4-d,\left[
g^{0}\right] =\frac{4-d}{2},
\end{equation}
suggesting that the fixed point may be accessible within the standard epsilon expansion in $d=4-\epsilon$ dimensions spacetime. Here we write the couplings with a superscript to emphasize that these are bare values rather than renormalized values.

Three kinds of QED$_{3}$-GNY models are of interests for us. The first one is the chiral Ising model\cite{gnyfol}, given by
\begin{equation}
\mathcal{L}_{\psi \phi,I}=-g\bar{\psi}_{i}\phi \psi _{i},
\end{equation}
it includes a real single component scale field $\phi$ ($N_b=1$). The chiral Ising model have undergone extensive study in recent times\cite{gnythl,gnyfol,edgny2,edgny3,edgny4}, for $N_f=1$, it has been suggested as a dual description of NC$\mathbb{C}$P$^1$ model, which describes a deconfined QCP between Neel and VBS transition on square lattice\cite{edgny2}; for $N_f=2$, it has been suggested to describe criticality between algebraic spin liquid and a chiral spin liquid on a spin-1/2 kagome antiferromagnet. The second one is the chiral XY model, given by
\begin{equation}
\mathcal{L}_{\psi \phi,XY}=-g\bar{\psi}_{i}(\phi_1+i\gamma^5\phi_2)\psi _{i},
\end{equation}
where the matrix $\gamma^5$ arises from the definition of two-component complex order parameter. Applications of the chiral XY model in the condensed matter context include semimetal-superconductor quantum criticality in graphene and topological insulators\cite{sccri}, Kekule VBS transition in graphene\cite{elefra}, and the emergent supersymmetric critical point at the boundary of a topological phase with $N_f=2$\cite{stsup}. Another QED$_{3}$-GNY model, which breaks spin-rotational symmetry spontaneously, refers to chiral Heisenberg model, and the Yukawa coupling is given as
\begin{equation}
\mathcal{L}_{\psi \phi,H}=-g\bar{\psi}_{i}[\vec{\sigma}\otimes1_{2N_f}\cdot\vec{\phi}] \psi _{i},
\end{equation}
where the Heisenberg order parameter $\vec{\phi}$ is a three component vector, $\vec{\sigma}$ denotes three Pauli matrix. The critical exponents of the GNY model has been studied at four loops, see for example Ref~[\onlinecite{gnyfol}]. However, the gauged chiral XY QED$_{3}$-GNY model and gauged chiral Heisenberg QED$_{3}$-GNY model, to the best of our knowledge, have rare been studied. Here, as will be shown below, the chiral QED$_{3}$-GNY model with an arbitrary $N_f$ and $N_b$ can be studied in an unified way, as only the commutator and the dimensions of the gamma matrices are required to reach the last results.

\subsection{Feynmann rules}
For the general chiral QED$_{3}$-GNY model, the Feynman rules can be read off directly from the Lagrangian Eq.(\ref{e1}) to Eq.(\ref{e3}). The propagator for fermion, scalar field and gauge field, respectively, are given by
\begin{align}
\psi: iS^0_{\alpha\beta}(p)&=i\delta_{\alpha\beta}(p\!\!\!/-M+i\epsilon)^{-1},\\
\phi: iD^0_{ab}(p)&=i\delta_{ab}(p^2-m^2+i\epsilon)^{-1},\\
A: iG^0_{\mu\nu}(k)&=-i\frac{g_{\mu\nu}-(1-\xi)k_{\mu}k_{\nu}/k^2}{k^2+i\epsilon}.
\end{align}
The 0 superscript here implies these are bare propagators. The fermion and scalar field propagators are given for the massive case, the massless propagator are obtained by setting $M=m^2=0$. In addition to the propagator, there are three vertices. The QED$_{3}$ and Yukawa vertex are given by $[\gamma_{A\bar{\psi}\psi}]^{\mu}=-ie\gamma^{\mu}$, $[\gamma_{\phi\bar{\psi}\psi}]^a=-ig\Sigma^a$, respectively. With this notation, the  QED$_{3}$ and Yukawa vertex functions, at lowest order, take the form
\begin{align}\label{vertexf}
  -ie\Gamma_{A\psi\psi}^{\mu}(p',p)=-ie\gamma^{\mu},\\
  -ig\Gamma_{\phi\psi\psi}^{a}(p',p)=-ig\Sigma^a.
\end{align}
Finally, the bosonic self-interaction vertex reads
\begin{equation}
  \gamma_{ijmn}=-i(\lambda/3)(\delta_{ij}\delta_{mn}+\delta_{im}\delta_{jn}+\delta_{in}\delta_{jm}).
\end{equation}
The lowest order bosonic vertex function $V_{ijmn}$ takes the form $iV_{ijmn}=\gamma_{ijmn}$.

\section{Renormalization group analysis and techniques}\label{section3}
To perform standard field-theoretic RG analysis in $d=4-\epsilon $ dimensions spacetime, we first define the bare theory.
The bare Lagrangian for the general chiral QED$_{3}$-GNY model is defined by Eq.(\ref{e1}) to (\ref{e3}) but with the fields and couplings are replaced by their bare counterparts,
$A\rightarrow A_{0}$, $\psi\rightarrow \psi_{0}$, $\phi\rightarrow \phi_{0}$, $e\rightarrow e_{0}$, $\lambda\rightarrow \lambda_{0}$,
$g\rightarrow g_{0}$, $m\rightarrow m_{0}$, $\xi\rightarrow \xi_{0}$.
Then, the renormalized Lagrangian is given by
\begin{align}\label{renol}
\mathcal{L}_R &=Z_{\psi}\bar{\psi}_ii\partial\!\!\!/_{\mu}\psi_i-\frac14Z_AF_{\mu\nu}^{2}-\frac{1}{2\xi}(\partial_{\mu}A_{\mu})^2  \notag \\
&-e\mu^{\epsilon/2}Z_{\psi }Z_A^{1/2}Z_e\bar{\psi}_{i}A\!\!\!/_{\mu}\psi_i+\frac{1}2Z_{\phi}(\partial_{\mu}\phi _{a})^2  \notag \\
&-\frac{1}{2}Z_{\phi }Z_{m^2}m^2\phi_a^{2}-\frac{\lambda }{4!}\mu^{\epsilon }Z_{\lambda}Z_{\phi }^{2}\left(\phi_a\phi
_{a}\right)^2  \notag \\
&-g\mu^{\epsilon /2}Z_{\psi}Z_g\sqrt{Z_{\phi}}\bar{\psi}_i\left(\Sigma^a\phi _a\right) \psi_i,
\end{align}
where $\mu $ denotes the energy scale parameterizing the RG flow, the explicit energy scale dependencies in the Lagrangian arise from the introduction of dimensionless couplings,
$g\mapsto g\mu ^{\epsilon /2}$, $\lambda \mapsto\lambda \mu ^{\epsilon }$ and $e\mapsto e\mu ^{\epsilon /2}$.
We have also defined the fields renormalization constants $Z_{\psi}$, $Z_{\phi}$ and $Z_{A}$, such that $\psi _{0}=\sqrt{Z_{\psi }}\psi $,
$\phi _{0}=\sqrt{Z_{\phi }}\phi$ and $A_{0\mu }=\sqrt{Z_{A}}A_{\mu }$.
The renormalization relate the bare and renormalized values as
\begin{equation}
X_{0}=Z_{X}X,\quad X=e,\lambda, g, m^2.
\end{equation}
Further, we have also defined the relation $\xi _{0}=Z_{A}\xi $ for the demand that the gauge-fixing term is 
form-invariant.
In the minimal subtraction scheme, the renormalization constants depend only on the dimensionless couplings and $1/\epsilon $ and can be
expanded into formal Laurent series, yielding
\begin{equation}
Z_X(\epsilon ,\lambda ,g,e)=1+\sum_{k=1}^{\infty }\frac{Z_{X,k}(\lambda,g,e)}{\epsilon^k}.
\end{equation}
The beta functions and the anomalous dimensions can be derived from these
renormalization constants. By employing the dimensional regularization, we evaluate these renormalization constants in Appendix \ref{appendixb} with minimal subtraction.

\subsection{Beta functions}

The beta function are defined as the logarithmic derivatives with respect to the energy scale $\mu $,
\begin{equation}
\beta _{e}=\frac{de}{d\ln \mu },\beta _{g}=\frac{dg}{d\ln \mu },\beta
_{\lambda }=\frac{d\lambda }{d\ln \mu }, \label{betaf}
\end{equation}
where the coupling constants $e$, $g$ and $\lambda $ are all dimensionless.
Since the bare couplings are independent of $\mu$, the beta function can be written as
\begin{equation}
\beta _{X}=-X\frac{d\ln Z_{X}}{\ln \mu },\quad X=e,\lambda ,g.
\end{equation}
Here, the renormalization constants are displayed in Appendix \ref{appendixa}.
At the leading order, the beta function for the gauge charge reads
\begin{equation}\label{betae}
\beta _{e}=-\frac{\epsilon }{2}e+\frac{N_{f}e^{3}}{12\pi ^{2}},
\end{equation}
this equation takes the same form as the four-component spinor quantum electrodynamics in $3+1$ dimensional spcetime.
The beta function for the Yukawa coupling is given by
\begin{equation} \label{betag}
\beta _{g}=-\frac{\epsilon}2g-\frac{6ge^2}{(4\pi)^{2}}+\frac{\left(
2N_{f}-N_{b}+4\right)g^3}{(4\pi)^2}.
\end{equation}
Furthermore, the beta function for the quartic scalar coupling is given by
\begin{equation} \label{betalm}
\beta_{\lambda}=-\epsilon \lambda +\frac{8N_fg^2\lambda }{(4\pi )^2}+\frac{\lambda^2(N_b+8)/3}{(4\pi )^2}-\frac{48N_{f}g^4}{(4\pi )^2}.
\end{equation}
Following the notation of Ref.~[\onlinecite{edgny2}], rescaling the couplings according to $\alpha^{2}/(8\pi ^{2})\mapsto \alpha ^{2}$, $\alpha ^{2}=e^{2}$, $g^{2}$, $\lambda $, and replacing the quartic scalar coupling according to $\lambda/4!\mapsto \lambda $,
our beta functions fully agree with the expressions in the chiral Ising QED$_{3}$-GNY model\cite{edgny3,edgny4}. 
The consistency serves as a nontrivial check on our calculations. In the pure QED$_3$ limit, $\beta_e$ recovers the well-known QED$_{3}$ beta function.
Setting $e=0$ and $g=0$, $\beta_{\lambda}$ agrees exactly with the O($N_b$) scalar field theory.
In the GNY limit, upon setting $e=0$,
$\beta_g$ and $\beta_{\lambda}$ are consistent with the beta function of GNY model\cite{gnyfol}.

\subsection{Anomalous dimensions}

The anomalous dimensions is defined as the logarithmic derivatives of the
renormalization constants with respect to the energy scale $\mu $,
\begin{equation}\label{anomalous}
\eta_X=\frac{1}{Z_X}\frac{dZ_X}{d\ln \mu }=\sum_{X=e,g,\lambda }\frac{d\ln Z_X}{dX}\beta_X
\end{equation}
where $Z_{X}\in \{Z_{\psi},Z_{\phi},Z_A,Z_{m^2}\}$ are renormalization constants and given in Appendix \ref{appendixa}.
At the critical point, the anomalous dimensions govern the scaling behavior and thus are universal.
Explicitly, the one-loop anomalous exponent of the scalar field and fermionic field are given by
\begin{align}
\eta_{\phi }&=\frac{4N_{f}g^{2}}{(4\pi )^{2}}, \label{anomphi} \\
\eta_{\psi}&=\frac{N_bg^2-(4-d-2\xi )e^2}{(4\pi)^2}.
\end{align}
Note that $\eta_{\psi }$ is gauge dependent. However, $\eta_{\phi }$ has no dependence on the gauge-fixing parameter, it governs the power law of the two-point correlation function at the critical point.
Using Eq.~(\ref{anomalous}), one gets the anomalous dimension for the mass square
\begin{equation}
\eta_{m^2}=-\frac{\lambda(N_b+2)/3+4N_fg^2}{(4\pi)^2}. \label{anomm}
\end{equation}
Our expressions can be checked in different limit.
Upon setting $e=0$ only, our results fully agree with the GNY model\cite{gnyfol}.
In the pure QED$_3$ limit, we recover the known result $\eta_{\psi}=2e^2/(4\pi)^2$ in the Feynman gauge.
Upon setting $g=e=0$, we have $\eta_{m^2}=-(N_b+2)\lambda/(48\pi^2)$ for the $O(N_b)$ scalar field theory.
\subsection{Nonsinglet fermion bilinears operators}
The mass term in the chiral QED$_{3}$-GNY model is a singlet fermion bilinear $\bar{\psi}\psi$ which is gauge invariant. Another gauge invariant symmetry-breaking operator we can consider is the non-singlet fermion bilinear.
Explicitly, we consider the QED$_{3}$--Ising-Yukawa theory with single flavor ($N_f=1$) four-component Dirac spinor, 
\begin{equation}
\mathcal{L}=\bar{\psi}\gamma^{\mu }(i\partial_{\mu }-eA_{\mu })\psi-g\phi\bar{\psi}\psi,
\end{equation}
where the $4\times4$ gamma matrices $\gamma^{\mu}$ form a four dimensional representation of Clifford algebra.
Rewriting the four-component spinor as $\psi=(\psi_1,\psi_2)$ and defining $\gamma^{\mu}=\sigma^z\otimes\tilde{\gamma}^{\mu}$, where the $2\times2$ gamma matrices $\tilde{\gamma}^{\mu}$ form a two dimensional representation of Clifford algebra.
The model can be expressed as 
\begin{equation}
  \mathcal{L}=\sum_{i=1}^2\bar{\psi}_i\tilde{\gamma}^{\mu }(i\partial_{\mu }-eA_{\mu })\psi_i-g\phi(\bar{\psi}_1\psi_1-\bar{\psi}_2\psi_2),
\end{equation}
with $\bar{\psi}_i=\psi^{\dag}_i\tilde{\gamma}^0$, $i=1,2$. 
The SU($2$) flavor symmetry is broken down to SU($1$)$\times$ SU($1$) owing to the presence of nonsinglet fermion bilinear $\bar{\psi}\sigma^z\psi$. In general, the flavor-symmetry breaking non-singlet SU($N_f$) fermion bilinear is defined as $\bar{\Psi}T_A\Psi$, where the generators $T_A$ of SU($N_f$) is a traceless $N_f\times N_f$ Hermitian matrix.
In algebra spin liquid, the non-singlet fermion bilinear bridge between field operators and physical observables like Neel and VBS order\cite{aspher}. For the conjectured QED$_3$-GNY--$\mathbb{C}$P$^1$ duality, the scaling dimensions of the non-singlet fermion bilinear is essential to establish the duality\cite{dw1,edgny3,edgny4}.
Here, we present some technics for reaching the scaling dimensions of nonsinglet fermion bilinears.

\begin{figure}[tbp]
\centering
\scalebox{0.9}{\includegraphics*{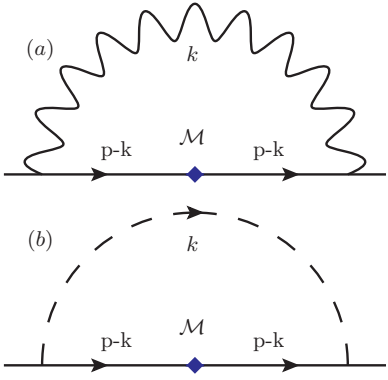}}
\caption{(Color online) One-loop corrections to the nonsinglet fermion bilinear $\mathcal{M}\bar{\psi}T_A\psi$.}
\label{fig1}
\end{figure}

The scaling dimensions of SU($N_f$) bilinear can be determined with the help of nonsinglet fermion bilinear insertions. More precisely, we add the renormalized SU($N_f$) fermion bilinear,
$\delta\mathcal{L}_R=Z_{\mathcal{M}}Z_{\psi}\mathcal{M}\bar{\psi}T_A\psi$, to the renormalized Lagrangian and then calculate the one-loop correction (see Fig.~\ref{fig1}). Here $\mathcal{M}$ serves as an infinitesimal background field that couples to the nonsinglet fermion bilinear and we have introduced renormaliztion constant $Z_{\mathcal{M}}$, which implies the anomalous dimension
$\eta_{\mathcal{M}}=d\ln Z_{\mathcal{M}}/d\ln\mu$. In this case, the nonsinglet fermion bilinear scaling dimension at a non-Gaussian fixed point reads
\begin{equation} \label{fbilinear}
\Delta_{\bar{\psi}T_A\psi }=d-1-\eta_{\mathcal{M}},
\end{equation}
where $\eta_{\mathcal{M}}$ is the anomalous dimension evaluated at the fixed point $\eta_{\mathcal{M}}\equiv
\eta_{\mathcal{M}}(e_{\ast}^{2},g_{\ast}^{2},\lambda_{\ast})$.
Since the fermion bilinear are gauge-invariant operators, $\eta_{\mathcal{M}}$ must be gauge invariant.
Fig.~\ref{fig1} shows the one-loop correction to the two-point function with infinitesimal $\mathcal{M}$ insertion.
To obtain $Z_{\mathcal{M}}$, we need to evaluate the one-loop corrections and determine the fermion bilinear counter term.

The Feynman diagram in Fig.~\ref{fig1}(a) gives
\begin{align}
-i\mathcal{M}&\Sigma^{c}\Gamma^{(1)}_{\bar{\psi}T_A\psi}(p)=\int \frac{d^{d}k}{(2\pi)^{d}}
\frac{1}{(p-k)^4}(-ie\gamma^{\mu}){i(p\!\!\!/-k\!\!\!/)}\notag\\
&\times(-i\mathcal{M}\Sigma^{c}){i(p\!\!\!/-k\!\!\!/)}(-ie\gamma^{\nu})G_{\mu\nu}(k),\notag \\
&=:-ie^2\mathcal{M}\Sigma^c\frac{8-2(1-\xi)}{(4\pi)^2\epsilon}.
\end{align}
To reach the result we have set the external momentum to be zero. The Fig.~\ref{fig1}(b) gives
\begin{align}
&-i\mathcal{M}\Sigma^{c}\Gamma^{(2)}_{\bar{\psi}T_A\psi}(0)=\int \frac{d^{d}k}{(2\pi)^{d}}
\frac{i\delta_{ab}}{k^4(k^2-m^2)}(-ig\Sigma^{a}){ik\!\!\!/}\notag\\
&\times(-i\mathcal{M}\Sigma^{c}){ik\!\!\!/}(-ig\Sigma^{b})=:ig^2\mathcal{M}\Sigma^{c}\frac{2(2-N_b)}{(4\pi)^2\epsilon}.
\end{align}
The loop corrections contain divergent terms and need renormalization. A counter term should be added to absorb the infinite. Using the minimal subtraction, we have the renormalization condition:
\begin{align}
-i\Gamma^{(1)}_{\bar{\psi}T_A\psi}(0)-i\Gamma^{(2)}_{\bar{\psi}T_A\psi}(0)-i(Z_{\mathcal{M}}Z_{\psi}-1)=0.
\end{align}
To calculate the renormaliztion constant at one-loop, we expand the $Z$-factors as $Z_{\mathcal{M}}=1+{\delta}_{\mathcal{M}}$, $Z_{\psi}=1+{\delta}_{\psi}$. Taking into account the subtraction condition,
one has
\begin{align}
\delta_{\mathcal{M}}&=-\Gamma^{(1)}_{\bar{\psi}T_A\psi}(0)-\Gamma^{(2)}_{\bar{\psi}T_A\psi}(0)-\delta_{\psi}\notag\\
&=\frac{(5-2N_b)g^2-6e^2}{(4\pi)^2}\frac{1}{\epsilon},
\end{align}
and the anomalous dimensions $\eta_{\mathcal{M}}$ is given by
\begin{align}\label{etam}
\eta_{\mathcal{M}}=\frac{1}{Z_{\mathcal{M}}}\sum_{X=e,g}\frac{dZ_{\mathcal{M}}}{dX}\beta _{X}=\frac{6e^2-(5-2N_b)g^2}{(4\pi )^2}.
\end{align}
\begin{figure*}[tbp]
\centering
\scalebox{0.6}{\includegraphics*{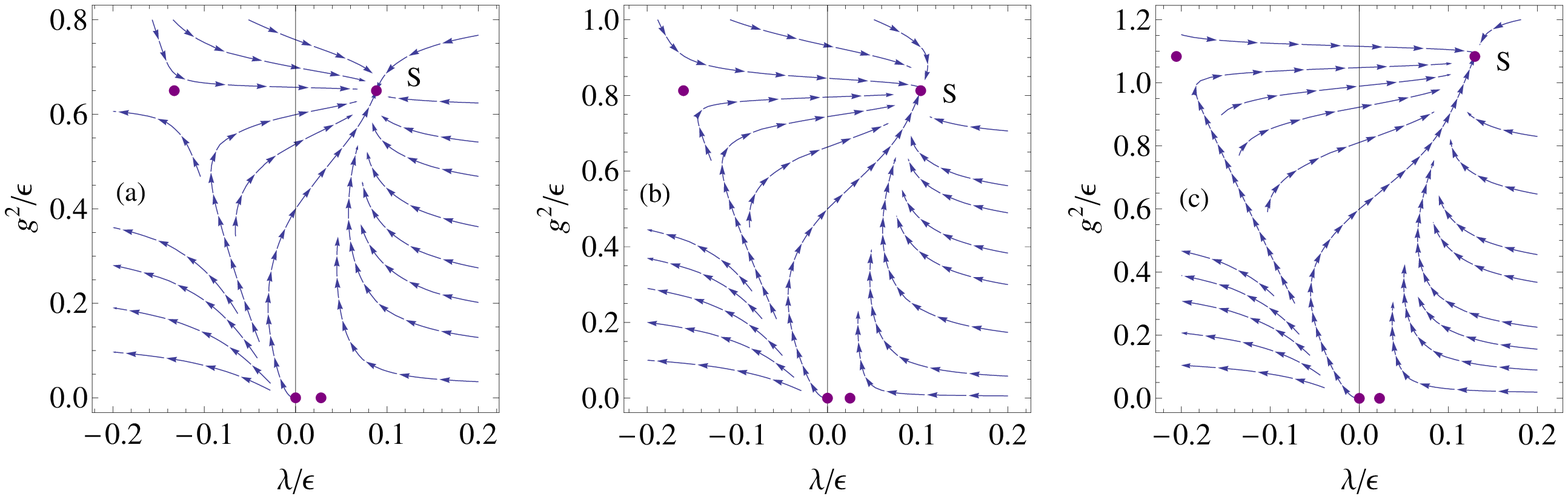}}
\caption{(Color online) The fixed-point structure and the RG flow of QED$_{3}$-GNY model in
$(g^2/(8\pi^2),\lambda/(8\pi^2))$ plane to leading order in $d=4-\epsilon$ dimensions. Here charge is chosen as its infrared stable QED fixed point $e^2=6\pi^2\epsilon/N_f$. (a)-(c) display the RG flow of chiral Ising, XY, and O(3) QED$_3$-GNY model, respectively. The critical behavior is governed by the infrared stable fixed point locates at S with $g^2>0$ and $\lambda>0$.}
\label{FS}
\end{figure*}
\section{Critical exponents}\label{section4}
To determine the critical exponents, we begin by searching for the fixed points with couplings
$(e_{\ast}^2,g_{\ast}^2,\lambda_{\ast})$. The beta functions Eqs.(\ref{betae})-(\ref{betalm}) admit eight fixed points: the Gaussian fixed point (0,0,0), the conformal QED$_3$ fixed point $(6\pi^2\epsilon/N_f,0,0)$,
the Wilson-Fisher fixed point $(0,0,48\pi^2\epsilon/(N_b+8))$, the QED$_3$--Wilson-Fisher fixed point with $e^2_{\ast}\neq 0$, $g_{\ast}^2=0$, $\lambda_{\ast}\neq0$. And two GNY fixed points with
$e^2_{\ast}=0$, $g_{\ast}^2\neq0$, $\lambda_{\ast}\neq0$.
Besides, there are two QED$_3$-GNY fixed points:
\begin{align}
e_{\ast}^2&=\frac{6}{N_f}\pi^2\epsilon,\\
g_{\ast}^2&=\frac{4(2N_f+9)}{N_f(4-N_b+2N_f)}\pi^2\epsilon, \\
\lambda_{\ast}^{\pm}&=24\frac{-(2N_f+N_b+14)\pm F}{(4-N_b+2N_f)(N_b+8)}\pi^2\epsilon,
\end{align}
with
\begin{align}
F=\sqrt{(2N_f+N_b+14)^2+4(N_b+8)(2N_f+9)^2/N_f}.
\end{align}
In agreement with previous study Ref.[\onlinecite{edgny2}], of all these fixed points only the positive QED$_3$-GNY fixed points $(e_{\ast}^2,g_{\ast}^2,\lambda^{+}_{\ast})$ are stable.
The fixed-point and the RG flow are shown in Fig.\ref{FS}, the positive QED$_3$-GNY fixed points corresponds to a continuous phase transition at which the symmetry is spontaneously broken. 
We also check that the positive stable fixed point established here holds only for $2N_f-N_b+4>0$, otherwise, one has a negative stable fixed point.

To study the universal critical behavior, we evaluate the fermion and boson anomalous dimension
\begin{equation}
\eta_{X}^{\ast }\equiv \eta _{X}(e_{\ast }^{2},g_{\ast }^{2},\lambda _{\ast}^{+}),
\end{equation}
where $X=\psi ,\phi ,A,m^{2}$. At the conformal QED$_3$ fixed point, the gauge
field anomalous dimension, to leading order in $d=4-\epsilon $, reads $\eta _{A}=\epsilon +\mathcal{O}(\epsilon ^{2})$, which arises from the Ward identity associated with the gauge U(1) symmetry.
The beta function for the mass square is conventionally defined as
$\beta_{m^2}=dm^2/d\ln{\mu}$, which yields
\begin{equation}
\beta_{m^2}=-(2+\eta _{m^2})m^2,
\end{equation}
In terms of the anomalous dimension of mass square at the fixed point, the inverse correlation length exponent $\nu^{-1}$ is obtained as
\begin{equation}\label{inverse}
\nu^{-1}=-d\beta _{m^2}/dm^2|_{\eta_{X}^{\ast }}=2+\eta_{m^2}^{\ast}.
\end{equation}
We are interesting with the exponent $\nu^{-1}$ and the scalar field anomalous anomalous dimension $\eta_{\phi}$. Let's evaluate the analytical expressions for the critical exponents of general QED$_3$-GNY theory.
At the QED$_3$-GNY fixed point, from Eqs.~(\ref{anomphi}), (\ref{anomm}) and (\ref{inverse}), we find
\begin{align}
\eta_{\phi}=\frac{2N_f+9}{(2N_f-N_b+4)}\epsilon+\mathcal{O}(\epsilon^2),
\end{align}
\begin{align}
{\nu}^{-1}&=2-\frac{(F-2N_f-N_b-14)(N_b+2)}{2(2N_f-N_b+4)(N_b+8)}\epsilon\notag\\
&-\frac{2N_f+9}{2N_f-N_b+4}\epsilon+\mathcal{O}(\epsilon^2).
\end{align}
The exponents of the chiral QED$_3$-GNY universality class can be easily calculated from the analytical expressions.
As a nontrivial check, we will see the above expressions recover the result of chiral Ising-QED$_3$--GNY universality class at corresponding order\cite{edgny3,edgny4}. In the following, we also calculate the exponents in chiral XY- and Heisenberg- QED$_3$-GNY universality class. As far as we are aware, the chiral QED$_3$-GNY universality class, especially the chiral XY QED$_3$-GNY and chiral O(3) QED$_3$-GNY universality class, have rare been studied.

\subsection{Chiral Ising QED$_{3}$-GNY universality calss}
In this section, we present the exponents of the intensely studied chiral Ising QED$_{3}$-GNY universality class recently.
The ungauged chiral Ising QED$_{3}$-GNY model with $N_f=2$ in graphene lattice has been argued to describe critical point of semimetal-CDW transition\cite{semicdw}.
The spinless fermions on two dimensional lattice with tuning repulsive interactions also belongs to the chiral Ising universality class\cite{gnythl,gnyfol}.  The ungauged model with $N_f=1/4$ has been argued to relevant to emergent supersymmetric critical point on the surface of topological phase\cite{stsup1}.
Furthermore, the chiral Ising QED$_{3}$-GNY model is relevant to the conjectured duality with the NC$\mathbb{C}$P$^1$ model and the quantum criticality for the spin-1/2 Kagome antiferromagnet\cite{edgny3,edgny4}.

When the system is tuned to criticality (Fig.\ref{FS}(a)), one gets the exponents as the function of fermion flavors $N_f$,
\begin{align}
\eta_{\phi}&=\frac{2N_f+9}{2N_f+3}\epsilon+\mathcal{O}(\epsilon^2),\\
{\nu}^{-1}&=2-\frac{10N_f^2+39N_f+W}{6N_f(2N_f+3)}\epsilon+\mathcal{O}(\epsilon^2),
\end{align}
with $W=\sqrt{4N_f^4+204N_f^3+1521N_f^{2}+2916N_f}$,
which agrees rather well with Ref.[\onlinecite{edgny3,edgny4}].
The $N_f=1$ case relevant for the conjectured duality between chiral Ising QED$_3$-GNY model and non-compact $\mathbb{C}$P$^1$ model, the numerical exponents give $\eta_{\phi}=2.2\epsilon$, $\nu^{-1}=2-3.90514\epsilon$.
For $N_f=2$, we obtain $\eta_{\phi}=1.85714\epsilon$, $\nu^{-1}=2-2.7937\epsilon$.
The plain extrapolation by setting $\epsilon=1$ gives an unphysical result, implying that the gauge field tend to worse the convergence of the $\epsilon$ expansion and change the ungauged chiral Ising GNY universality class qualitatively\cite{gnyfol}.
To obtain an estimates of critical exponents for the physical dimensions,
we employ simple Pad$\acute{e}$ approximants\cite{edgny3}.
At the lowest order, the Pad$\acute{e}$ approximants give an unique [0/1] extrapolation.

The pad$\acute{e}$ estimates for the order parameter anomalous dimensions and the inverse correlation-length exponent are listed in Table \ref{tab1}, we list the extensively study $N_f=1$ and $N_f=2$ case. Different Pad$\acute{e}$ approximants for the inverse correction length exponent $\nu^{-1}$ in the chiral Ising QED$_3$-GNY model\cite{edgny3,edgny4}, however,
have a rather large interval ranging from 0.0486 to 0.677.
The $\mathcal{O}(1/N)$ result seems an order of magnitude smaller than the $4-\epsilon$ expansion.
Therefore, it will be an inevitable task to give an improved estimates in future study.

Further, the ungauged chiral Ising QED$_3$-GNY model with $N_f=1/4$ is compatible with an emergent supersymmetric criticality at which the bosons and fermions are identical to each other. The critical exponents give
\begin{align}
\eta=\eta_{\phi}=\eta_{\psi}=\epsilon/7.
\end{align}
At one-loop order, we observe the superscaling relation $\nu^{-1}=(d-\eta)/2$ exactly holds\cite{supjhep}.
In general, the supersymmetric criticality always satisfied for $N_b=4N_f$, so the supersymmetric scenario also exists in the O(2) and O(4) universality class.
Finally, in the $e^2=g^2=0$ limit, the chiral QED$_3$-GNY model reduces to the completely decoupled chiral fermion theory and scalar $O(N_b)$ field theory.
And the critical behavior is dominated by the Wilson-Fisher fixed point, one then gets
\begin{align}
  \nu^{-1}=2-\frac{N_b+2}{N_b+8}\epsilon+\mathcal{O}(\epsilon^2), \quad \eta_{\phi}=0,
\end{align}
at leading order. Further, all the six critical exponents satisfy the scaling relations:
\begin{align}
\alpha&=2-d\nu, \beta=\nu(d-2+\eta)/2, \\
\gamma&=\nu(2-\eta),\gamma=\beta(\delta-1).
\end{align}

\begin{table}[tbp]
\caption{Critical exponents for chiral QED$_{3}$-GNY universality class in $d=2+1$ dimensional spacetime:  Estimates of inverse correlation length exponent $\nu^{-1}$ and fermions anomalous dimensions $\eta_{\phi}$ with [0/1] extrapolation at leading $\epsilon$ expansion.
For comparison, we also list the estimates from  $1/N$-expansion at $O(1/N)$ order.}
\begin{tabular}{llll}
\hline\hline
chiral Ising QED$_3$-GNY &$N_f$\hspace{0.4cm} &$\nu^{-1}$\hspace{1cm} &$\eta_{\phi}$\hspace{1cm} \\
\hline
this work $4-\epsilon$, [0/1]                      &1      &0.6774        &2.2        \\
$4-\epsilon$, [1/1],four-loop[\onlinecite{edgny3}] &1      &0.660         &2.00       \\
$4-\epsilon$, [1/1],three-loop[\onlinecite{edgny4}]&1      &0.6595        &1.9978      \\
$\mathcal{O}(1/N)$,[0/1],[\onlinecite{nexpan}]     &1      &...           &0.093       \\
this work $4-\epsilon$, [0/1]                      &2      &0.8344        &1.8571      \\
$\mathcal{O}(1/N)$,[0/1],[\onlinecite{nexpan}]     &2      &...           &0.17         \\
this work $4-\epsilon$, [0/1]                      &$\infty$ &1.3333        &1.0      \\
\hline
chiral XY QED$_3$-GNY  &$N_f$\hspace{0.4cm} &$\nu^{-1}$\hspace{1cm} &$\eta_{\phi}$\hspace{1cm} \\
this work $4-\epsilon$, [0/1]                      &1      &0.5374       &2.75         \\
$O(1/N)$, [0/1][\onlinecite{vbslqe}]               &1      &0.27         &...          \\
this work $4-\epsilon$, [0/1]                      &2      &0.7208       &2.1667        \\
$O(1/N)$ [0/1][\onlinecite{vbslqe}]                &2      &0.426        &...           \\
this work $4-\epsilon$, [0/1]                      &$\infty$ &1.3333        &1.0      \\
\hline
chiral O(3) QED$_3$-GNY  &$N_f$\hspace{0.4cm} &$\nu^{-1}$\hspace{1cm} &$\eta_{\phi}$\hspace{1cm} \\
this work $4-\epsilon$ [0/1]                       &1      &0.4024      &3.6667           \\
this work $4-\epsilon$, [0/1]                      &2      &0.6082      &2.6             \\
this work $4-\epsilon$, [0/1]                      &$\infty$ &1.3333        &1.0      \\
\hline\hline
\end{tabular}
\label{tab1}
\end{table}

\subsection{Chiral XY QED$_{3}$-GNY universality class }
In condensate condensed-matter, the chiral XY QED$_{3}$-GNY universality class is relevant to the quantum criticality at which the order parameter exhibits
U(1) symmetry, i.e., superconductivity criticality\cite{sccri} and semimetal-Kekule VBS transition in graphene \cite{elefra}.
At the stable chiral XY QED$_3$-GNY fixed point (Fig. \ref{FS}(b)), we get
\begin{align}
\eta_{\phi}&=\frac{2N_f+9}{2N_f+2}\epsilon+\mathcal{O}(\epsilon^2),\\
\nu ^{-1}&=2-\frac{8N_f^2+29N_f+W}{10N_f(N_f+1)}\epsilon+\mathcal{O}(\epsilon^2),
\end{align}
with $W=\sqrt{4N_f^4+224N_f^3+1696N_f^2+3240N_f}$.
For the correlation-length exponent we get $\nu^{-1}=2-5.44305\epsilon$ for $N_f=1$,
and $\nu^{-1}=2-3.54939\epsilon$ for $N_f=2$.
The result from pad$\acute{e}$ estimates of chiral XY QED$_{3}$-GNY universality class are listed in table \ref{tab1}, together with the estimate from $1/N$ expansion at $\mathcal{O}(1/N)$.
Again, we observe the criticality qualitatively changed in the ungauged chiral QED$_3$-GNY universality class.
\subsection{Chiral Heisenberg QED$_{3}$-GNY universality class}

The chiral Heisenberg QED$_{3}$-GNY  universality class refers to a theory with the order parameter field having three real components.
A possible candidate is the field-theoretical formulation of the Mott criticality of interacting electrons, i.e., the $N_f=2$ case corresponds to transition towards an antiferromagnetic state of interacting electrons on the honeycomb lattice\cite{qchub,unive,eplaf,prlaf}. At the stable QED$_{3}$-GNY fixed point (Fig.\ref{FS}(c)), one gets
\begin{align}
\eta _{\phi}&=\frac{2N_f+9}{2N_f+1}\epsilon+\mathcal{O}(\epsilon^2),\\
\nu ^{-1}&=2-\frac{34N_f^2+113N_f+5W}{22N_f(2N_f+1)}\epsilon+\mathcal{O}(\epsilon^2).
\end{align}
with $W=\sqrt{4N_f^4+244N_f^3+1873N_f^2+3564N_f}$.
The numerical evaluation of the correlation-length exponent provides $\nu^{-1}=2-7.93931\epsilon$ for $N_f=1$, and $\nu^{-1}=2-4.57683\epsilon$ for $N_f=2$.
In Table \ref{tab1}, we also provide the pad$\acute{e}$ estimate for the critical exponents in the chiral Heisenberg [or $O(3)$] QED$_3$-GNY universality class. When $N_f\rightarrow\infty$, the chiral Ising-, XY- and
$O(3)$-QED$_3$-GNY theory tend to a same universality class.

Finally, the critical exponents for the chiral QED$_3$-GNY model should be verified by other approaches, i.e., conformal bootstrap approach and sign-free quantum Monte Carlo simulations.
Recently, the Monte Carlo simulation of quantum rotor model pioneer
a new technique to study chiral QED$_3$-GNY universality class\cite{rotor}. The lattice quantum electrodynamics with interacting fermionic matter provide a platform for realizing various quantum phases and universality class\cite{algvbs}. By varying the coupling strength and the coupling constant of the gauge field, it's possible for us to reach the breaking phases such as N$\acute{e}$el antiferromagnet and valence-bond-solid. We therefore believe that the numerical results for the critical behavior of the continuum chiral QED$_3$-GNY model can be reached by the large-scale numerical Monte Carlo simulations.

\section{Fermion bilinears and conjectured duality relations}\label{section5}
We are interested in the scaling dimensions of SU($N_f$) flavor nonsinglet fermion bilinear $\Delta_{\bar{\psi}T_{A}\psi}$ in chiral QED$_{3}$-GNY model.
In chiral Ising QED$_{3}$-GNY universality class, the nonsinglet fermion bilinear relevant for the conjectured scaling relations (\ref{equal1}) and (\ref{equal2}) which bridge between the Ising QED$_{3}$-GNY model involving two flavors of two-component Dirac fermions andthe SU(2) NC$\mathbb{C}$P$^1$ model.
According to Eqs.(\ref{fbilinear}) and (\ref{etam}), at one-loop order, we obtain
\begin{align}
\Delta_{\bar{\psi}T_A\psi}=d-1-\frac{8N_f+9N_b+4N_bN_f-9}{4N_f(2N_f-N_b+4)}\epsilon,
\end{align}
where $d=4-\epsilon$. For $N_b=1$, this expression gives
\begin{align}
\Delta_{\bar{\psi}T_A\psi}=3-\frac{2N_f+6}{2N_f+3}\epsilon+\mathcal{O}(\epsilon^2),
\end{align}
which in agreement with Ref.[\onlinecite{edgny3}].
Strictly speaking, the flavor nonsinglet fermion bilinear only exists for SU($N_f$) with $N_f\geq 2$, implying the absence of flavors symmetry breaking at SU($1$) case. Estimates for the exponents in three dimensions depend on the radius of convergence of series $\mathcal{O}(\epsilon^2)$. To provide a comparable estimates, we have evaluated the SU($N_f$) fermion bilinear scaling dimensions for various $N_f$ using the Pad$\acute{e}$ approximants. Our estimates are shown in Table \ref{tab2}.
We observe that fermion bilinear scaling dimensions increase monotonically as $N_f$ increase in the chiral QED$_{3}$-GNY universality class, an interesting feature is that the scaling dimensions has a
maximum value as $N_f\rightarrow\infty$ in this universality class,
\begin{align}
\Delta^{max}_{\bar{\psi}T_A\psi}=2.5.
\end{align}

\begin{table}[tbp]
\caption{ Pad$\acute{e}$ estimates for the flavor nonsinglet fermion bilinear $\Delta_{\bar{\psi}T_{A}\psi}$ using $[0/1]$ extrapolation in the chiral QED$_{3}$-GNY universality class. The results quoted from $1/N$ expansion are also listed for comparison.}
\begin{tabular}{lll}
\hline\hline
chiral Ising QED$_3$-GNY\hspace{0.6cm} &[0/1]\hspace{1cm}\quad &$O(1/N^2)$[\onlinecite{biope}]\\
\hline
$N_f=1$   &1.9565 &1.3795\\
$N_f=2$   &2.0323 &1.6482\\
$N_f=4$   &2.1064 &1.8116\\
$N_f=10^5$ &2.2499 &1.9213\\
\hline
chiral XY QED$_3$-GNY\hspace{0.6cm} &[0/1]\hspace{1cm}\quad &$O(1/N^2)$\\
$N_f=1$   &1.6180 &...\\
$N_f=2$   &1.8541 &...\\
$N_f=4$   &2.0196 &...\\
$N_f=10^5$  &2.2499 &...\\
\hline
chiral O(3) QED$_3$-GNY\hspace{0.6cm} &[0/1]\hspace{1cm}\quad &$O(1/N^2)$\\
$N_f=1$   &1.2558 &...\\
$N_f=2$   &1.6514 &...\\
$N_f=4$   &1.9229 &...\\
$N_f=10^5$  &2.2499 &...\\
\hline\hline
\end{tabular}
\label{tab2}
\end{table}

We now discuss a application of the scaling dimensions of SU($N_f$) adjoint flavor nonsinglet fermion bilinear.
Building on early fascinating works, the two flavors of two-component Dirac fermions chiral QED$_{3}$-GNY model has recently been conjectured to be dual to the NC$\mathbb{C}$P$^1$ model\cite{dw1,edgny2}. If the duality holds, it would imply an emergent SO(5) symmetry in both models when they are tuned to their putative critical points.
An implication of the duality is the scalar anomalous dimension $\eta_{\phi}$ should coincide at criticality,
\begin{equation}\label{equal1}
\eta_{\text{VBS}}=\eta_{\text{Neel}}=\eta_{\phi},
\end{equation}
where the first equality arises from SO(5) symmetry on the NC$\mathbb{C}$P$^1$ side and the second one from $z^{\dag}\sigma^zz\sim\phi$.
Furthermore, both $z^{\dag}z$ and $\phi^2$ are all rank-2 tensors of the SO(5) vector\cite{dw1,edgny3,edgny4}, the former tune through Neel-VBS criticality in the NC$\mathbb{C}$P$^1$ model and the latter tune through transition in the chiral QED$_{3}$-GNY model. Therefore, the duality between rank-2 operators imply the correlation-length exponents in two models should coincide,
\begin{equation}
\nu_{\text{CP$^1$}}=\nu_{\text{QED$_3$-GNY}}.
\end{equation}
According to the proposed duality, the fermion bilinear $\bar{\psi}T_A\psi$ has the same correlations as $\phi^2$ in the chiral QED$_{3}$-GNY model, which implies the scaling relation
\begin{equation}\label{equal2}
  [\bar{\psi}T_A\psi]=3-\nu^{-1}_{\text{QED$_3$-GNY}}.
\end{equation}
This nontrivial scaling relation is particularly interesting as it only relates the scaling dimensions of different operators on the QED$_{3}$-GNY side, which allow us to check the conjectured duality on a quantitative level.
The present one-loop result gives
\begin{equation}
  3-\nu^{-1}=2.3226.
\end{equation}
which agrees within $16\%$ with the scaling dimensions of fermion bilinear. 
Numerical studies of N$\acute{e}$el-VBS transition in spin systems find $1/\nu$ ranging from $1.28$ to $2.0$\cite{cptn1,kaul,pujari,bartosch,nahum2},
these values are inconsistent with our estimate of $\nu^{-1}=0.6774$ for chiral Ising QED$_3$-GNY model.
The anomalous dimensions $\eta_{\text{N$\acute{e}$el}}$ and $\eta_{\text{VBS}}$ for the N$\acute{e}$el and VBS order parameters, according to the proposed duality, should coincide with $\eta_{\phi}$.
The numerical simulations yield $\eta_{\text{N$\acute{e}$el}}\approx\eta_{\text{VBS}}$\cite{pujari}, with the exponents given near $0.3$. Unfortunately, the one-loop result for $\eta_{\phi}$ gives $1.9$, which is roughly an order of magnitude larger than the exponent in bosonic systems. As far as we are aware, the inconsistency remains at higher-order perturbative calculations\cite{edgny3,edgny4}. So it will be an essential task to track the origin for that inconsistency in the future.

The one-loop result of the scaling dimensions of the SU($N_f$) adjoint flavor nonsinglet fermion bilinear predict the upper boundary $\Delta_{\bar{\psi}T_A\psi}<2.25$ in the chiral QED$_3$-GNY theory.  
If the conjectured strong Ising-QED$_3$-GNY--$\mathbb{C}$P$^1$ holds at the critical point, then the inverse correction length exponent should meet 
\begin{equation}
\nu^{-1}>0.75.
\end{equation}
In this case, we expect the higher-loop result provide a positive correction for $\nu^{-1}$ and the non-perturbative approaches could provide an improved estimates. In summary, we have predicted the lower boundary of $\nu^{-1}$ for strong Ising-QED$_3$-GNY--$\mathbb{C}$P$^1$ duality. 

\section{Conclusions} \label{section6}
We have determined the critical behavior of the chiral QED$_3$-GNY model within the one-loop $\epsilon$ expansion in $d=4-\epsilon$ dimensional spacetime.  By employing the dimensional regularization and minimal
subtraction, the corresponding renormalization group equation is established. 
In analogy to the special chiral Ising QED$_3$-GNY model\cite{edgny3}, we found the general QED$_3$-GNY model exhibits an unique positive infrared-stable fixed point at which the criticality is universal, and the positive infrared-stable fixed point exists for the internal components of order parameter less than or equal to 5. 
Further, we have calculated the critical exponents and scaling dimensions of flavor-symmetry breaking fermion bilinear for the chiral Ising-, XY- and Heisenberg- QED$_3$-GNY universality class. The models investigated in this paper are relevant to Mott criticality of interacting electrons interacting with a gauge degrees of freedom.

The emergent supersymmetric quantum criticality has been conjectured as conformal field theories in the ungauged QED$_3$-GNY model. In this case, we found the positive infrared-stable fixed point also holds for the internal components of order parameter less than or equal to 5 and the supersymmetry holds for $4N_f=N_b$. Therefore, the supersymmetry criticality should exist in the Ising, XY and O(4) universality class.

Most interestingly, the present one-loop result for the inverse correction length exponent is consistent within $16\%$ with the conjectured dual relation that from the Ising-QED$_3$-GNY--$\mathbb{C}$P$^1$ duality at the critical point\cite{dw1,edgny3,edgny4}. We found the flavor-symmetry breaking nonsinglet fermion bilinear has a scaling dimensions with upper boundary $\Delta_{\bar{\psi}T_A\psi}<2.25$ in the chiral QED$_3$-GNY model. If the strong Ising-QED$_3$-GNY--$\mathbb{C}$P$^1$ duality holds at criticality, the inverse correction length exponent meets $\nu^{-1}>0.75$, which should be numerically checked in a suitable lattice model.

\begin{acknowledgments}
This work is supported partly by the NSFC under No.11647111, No.11674062, No.11974053, and partly by the Scientific Research Foundation of Guizhou Universality under No.20175788. The Feynman graphs were drawn with jaxodraw.
\end{acknowledgments}

\begin{appendix}
\section{One-loop corrections}\label{appendixa}
\begin{figure}[tbp]
\centering
\scalebox{0.8}{\includegraphics{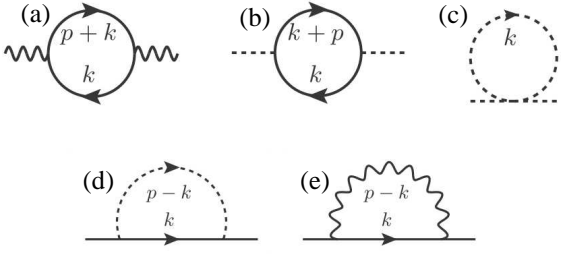}}
\caption{(Color online)The one-loop Feynman diagrams for self-energy. The wavy line indicates gauge propagator, the dot line indicates scalar fields propagator, the full line indicates fermion propagator. }
\label{self}
\end{figure}

In this appendix, we present the one-loop corrections to $n$-point function. The two-point function refers to self-energy correction, three three-point function refers to vertex correction, four-point function refers to bosonic self-interaction vertex.
Here, we work in $d=4-\epsilon$ dimensions, but ultimately we shall set $\epsilon=1$ to recover the physical dimensions in $d=3$ dimensions.
\subsection{Gauge self-energy}
The gauge field self-energy is diagrammatically represented by Fig.~\ref{self}(a), which gives
\begin{align}
i\Pi_{\mu \nu}&=(-1)N_f(-ie)^{2}\int \frac{d^dk}{(2\pi)^d}
\text{Tr}\left[\gamma ^{\mu }\frac{i}{k\!\!\!/}\gamma^{\nu }\frac{i}{(p\!\!\!/+k\!\!\!/)}\right] \notag \\
&=-N_{f}e^{2}\int \frac{d^dk}{(2\pi)^d}\text{Tr}\left[\gamma^{\mu}
\frac{k\!\!\!/}{k^2}\gamma^{\nu}\frac{(p\!\!\!/+k\!\!\!/)}{(p+k)^2}\right].
\end{align}
The minus sign $(-1)$ comes from the fermion loop factor.
Owing to the Lorentz invariance and Ward identity, it's convenient to extract the tensor structure form the gauge fields self-energy in the form
\begin{align}
  i\Pi_{\mu \nu}=(\delta ^{\mu \nu }p^{2}-p^{\mu }p^{\nu })\Pi(p),
\end{align}
where $\Pi(p)$ is a scalar function. Performing the integral over $k$ gives
\begin{align}
\Pi (p)=-i\frac{2N_{f}D_{\gamma }e^{2}}{3(4\pi )^{2}}\frac{1}{\epsilon}.
\end{align}
where $D_{\gamma}$ represents the matrix dimensions of Dirac gamma matrix.
Therefore, the divergent part of gauge fields self-energy reads
\begin{align}
i\Pi_{\mu \nu}=:-i(\delta ^{\mu \nu }p^{2}-p^{\mu }p^{\nu })\frac{2N_{f}D_{\gamma }e^{2}}{3(4\pi )^{2}}\frac{1}{\epsilon}.
\end{align}
Here and after in this paper, the symbol $=:$ means we only extract the divergent part when $d\rightarrow4$.
\subsection{Boson self-energy}
The Feynman diagrams in Fig.~\ref{self} (b) and (c) give the bosonic self-energy
$\Sigma_{\phi\phi}=\Sigma^{\psi}_{\phi\phi}+\Sigma^{\phi}_{\phi\phi}$. Fig.~\ref{self}(b) gives
\begin{align}
-i\Sigma^{\psi}_{\phi\phi}&=(-1)(ig)^{2}N_{f}\int \frac{d^dk}{(2\pi)^d}
\text{Tr}\left[\Sigma_a\frac{i}{k\!\!\!/}\Sigma _{b}\frac{i}{(k\!\!\!/+p\!\!\!/)
}\right] \notag \\
&=-g^2N_f\delta_{ab}\int \frac{d^dk}{(2\pi)^d}\text{tr}\left[ \frac{1}{k\!\!\!/}\frac{1}{(k\!\!\!/+p\!\!\!/)}\right].
\end{align}
where the minus sign $(-1)$ in the first row arises from the fermionic loop.
Evaluation of the integral gives
\begin{align}
-i\Sigma^{\psi}_{\phi\phi}=ig^{2}N_fD_{\gamma}\delta _{ab}\frac{p^2}{(4\pi )^2}\frac{1}{\epsilon}+\text{finite}.
\end{align}
Here, the finite part is rather complicated but
whose exact form does not concern us.
Diagrammatically, Fig.~\ref{self}(c) gives
\begin{align*}
-i\Sigma^{\phi}_{\phi\phi}&=\frac{1}{2}(-i\lambda/3)(\delta ^{ij}\delta
^{mn}+\delta ^{im}\delta ^{jn}+\delta ^{in}\delta ^{jm})\notag \\
&\times\int \frac{d^{d}k}{(2\pi)^d}\frac{i\delta ^{mn}}{k^2-m^2}=:i\lambda\frac{(N_{b}+2) m^{2}}{3(4\pi )^{2}}\frac{1}{\epsilon }\delta^{ij}.
\end{align*}
In terms of Dyson series, we have $D^{-1}=D^{-1}_0-\Sigma_{\phi\phi}$.
So the full Feynman propagator for scalar fields is given by
\begin{equation}
  iD_{ij}=i\delta_{ij}(p^2-m^2-\Sigma_{\phi\phi})^{-1}.
\end{equation}

\subsection{Fermion self-energy}
The Feynman diagrams in Fig.~\ref{self} (d) and (e) give the bosonic self-energy
$\Sigma_{\psi\psi}=\Sigma^{\phi}_{\psi\psi}+\Sigma^{A}_{\psi\psi}$, Fig.~\ref{self}(d) gives
\begin{align}
-i\Sigma_{\psi\psi}^{\phi}&=(-ig)^{2}\int \frac{d^dk}{(2\pi)^d}
\Sigma_a\frac{i}{k\!\!\!/}\Sigma_b\frac{i\delta_{ab}}{(p-k)^2-m^2}\notag \\
&=:ig^2\frac{N_b}{(4\pi)^2}\frac{p\!\!\!/}{\epsilon}.
\end{align}
Fig.~\ref{self}(e) gives
\begin{align}
-i\Sigma_{\psi\psi}^{A}&=(-ie)^{2}\int \frac{d^dk}{(2\pi)^d}
\gamma^{\mu}\frac{i}{k\!\!\!/}\gamma^{\nu}(-i)G_{\mu \nu }(p-k),\notag \\
&=I_1+(1-\xi)I_2.
\end{align}
where $G_{\mu\nu}(p-k)$ is the gauge propagator. $I_1$ is given by
\begin{align}
I_1&=-e^{2}\int \frac{d^dk}{(2\pi)^d}\gamma^{\mu}\frac{k\!\!\!/}{k^2}
\gamma^{\mu }\frac{1}{(p-k)^2} \notag \\
&=-e^{2}\int_{0}^{1}dx\int \frac{d^{d}k}{(2\pi )^{d}}\frac{(\epsilon-2)k\!\!\!/}{
[xk^{2}+(1-x)(p-k)^{2}]^2}\notag\\
&=:-(\epsilon-2)\frac{ie^{2}}{(4\pi)^{2}}\frac {p\!\!\!/}{\epsilon}.
\end{align}
where $\gamma ^{\mu }\gamma ^{\nu}\gamma ^{\mu }=-(2-\epsilon)\gamma ^{\nu }$ has been used.
$I_2$ is given by
\begin{align}
I_2=e^2\int \frac{d^dk}{(2\pi)^d}
(p\!\!\!/-k\!\!\!/)\frac{k\!\!\!/}{k^2}\frac{(p\!\!\!/-k\!\!\!/)}{(p-k)^{4}}=:\frac{-i2e^{2}}{(4\pi )^{2}}\frac{p\!\!\!/}{\epsilon}.
\end{align}
Specifically, $\xi=1$ corresponds to Feynman gauge and $\xi=0$ corresponds to Lorentz (covariant) gauge. In the Lorentz gauge,
$I_1$ cancels $I_2$ exactly as $d\rightarrow4$.
Now, in an arbitrary gauge, the full fermion self-energy reads
\begin{align}\label{fermionicselfa8}
-i\Sigma _{\psi\psi }=:i\frac{N_bg^2-(\epsilon-2)e^2-2(1-\xi)e^2}{(4\pi)^2}\frac{p\!\!\!/}{\epsilon}.
\end{align}

\begin{figure}[tbp]
\centering
\scalebox{0.8}{\includegraphics{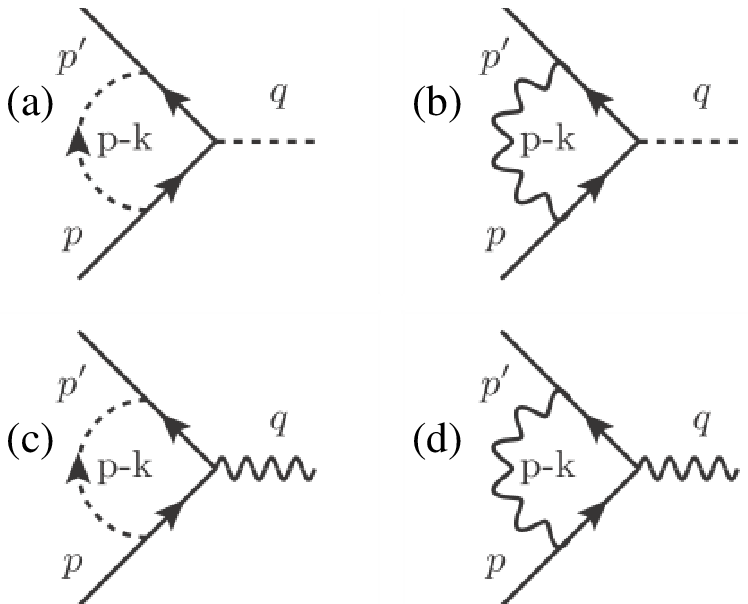}}
\caption{(Color online)The one-loop Feynman diagrams for vertex. The wavy line indicates gauge propagator, the dot line indicates scalar fields propagator, the full line indicates fermion propagator.}
\label{ver}
\end{figure}
\subsection{Yukawa vertex $\Gamma_{\phi\psi\psi}$ }
The fermions couple with gauge fields and scalar fields
in the Lagrangian. The Feynman diagrams Fig.~\ref{ver} gives the Yukawa vertex
$[\Gamma_{\phi\psi\psi}]^b(q)=[\Gamma^{\phi}_{\phi\psi\psi}]^b(q)+[\Gamma^{A}_{\phi\psi\psi}]^b(q)$.
Fig.~\ref{ver}(a) gives
\begin{align*}
-ig[\Gamma^{\phi}_{\phi\psi\psi}]^b(q)&=\int \frac{d^dk}{(2\pi )^d}\frac{ig^3\delta_{ac}}
{[(p-k)^{2}-m^{2}](q+k)^2k^2}\notag\\
&\times(-i\Sigma^a)i(q\!\!\!/+k\!\!\!/)(-i\Sigma^b)ik\!\!\!/(-i\Sigma^c).
\end{align*}
To eliminate divergence in the integral, it's convenient to calculate the vertex corrections with zero external momentum. Setting $q=p'=p=0$, one get
\begin{align}
-ig[\Gamma^{\phi}_{\phi\psi\psi}]^b(0)&=g^{3}\Sigma^b(2-N_{b})\int
\frac{d^dk}{(2\pi)^d}\frac{1}{k^2-m^2}\frac{1}{k^2} \notag \\
&=:ig^3\Sigma^{b}\frac{2(2-N_b)}{(4\pi)^{2}}\frac{1}{\epsilon}.
\end{align}
Similarly, Fig.~\ref{ver}(b) gives
\begin{align}
&-ig[\Gamma^{A}_{\phi\psi\psi}]^b(0)=e^2\int \frac{d^{d}k}{(2\pi)^d}
(-i)G_{\mu\nu}(-k)(-i\gamma^{\mu})\frac{i}{k\!\!\!/}\notag\\
&\times(-ig\Sigma ^b)\frac{i}{k\!\!\!/}(-i\gamma^{\nu})=I_1(0)+(1-\xi)I_2(0).
\end{align}
\begin{align}
I_1(0)&=e^2\int \frac{d^{d}k}{(2\pi )^d}\frac{
-i\delta _{\mu \nu }}{k^2}(-i\gamma ^{\mu })\frac{i}{k\!\!\!/}(-ig\Sigma^b)\frac{i}{k\!\!\!/}(-i\gamma^{\nu })\notag \\
&=:-i8ge^{2}\Sigma ^{b}\frac{1}{(4\pi )^{2}}\frac{1}{\epsilon}.
\end{align}
\begin{align}
I_2(0)&=e^2\int\frac{d^dk}{(2\pi )^{d}}\frac{
-i(-)(1-\xi)}{k^2}\frac{k_{\mu }k_{\nu}}{k^2}(-i\gamma^{\mu })\frac{i}{k\!\!\!/}\notag\\
&\times(-ig\Sigma^{b})\frac{i}{k\!\!\!/}(-i\gamma^{\nu})=:
ige^{2}\Sigma^{b}\frac{2}{(4\pi )^{2}}\frac{1}{\epsilon}.
\end{align}
Therefore, one get
\begin{align}
-ig[\Gamma^{A}_{\phi\psi\psi}]^b(0)=:-ige^{2}\Sigma^{b}\frac{8-2(1-\xi)}{(4\pi)^{2}}
\frac{1}{\epsilon }.
\end{align}

\subsection{QED$_3$ vertex $\Gamma_{A\psi\psi}$}
The Feynman diagrams in Fig.~\ref{ver} (c) and (d) give the Yukawa vertex
$[\Gamma_{A\psi\psi}]^{\mu}(q)=[\Gamma^{\phi}_{A\psi\psi}]^{\mu}(q)+[\Gamma^{A}_{A\psi\psi}]^{\mu}(q)$.
Setting $q=p'=p=0$, Fig.~\ref{ver}(c) gives
\begin{align}
&-ie[\Gamma^{\phi}_{A\psi\psi}]^{\mu}(0)=\int \frac{d^dk}{(2\pi)^d}\frac{i\delta
_{ac}(-ig\Sigma^a)ik\!\!\!/}{[k^2-m^2]k^4}(-ie\gamma^{\mu})ik\!\!\!/\notag \\
&\times(-ig\Sigma^c)=:-ieg^2N_b\gamma^{\mu }\frac{1}{(4\pi)^2}\frac{1}{\epsilon}.
\end{align}
Fig.~\ref{ver}(d) gives
\begin{align}
&-ie[\Gamma^{A}_{A\psi\psi}]^{\alpha}(0)=e^3\int \frac{d^dk}{(2\pi)^d}
(-i)G_{\mu\nu}(-k)(-i\gamma ^{\mu })\frac{i}{k\!\!\!/} \notag \\
&\times(-i\gamma^{\alpha })\frac{i}{k\!\!\!/}(-i\gamma ^{\nu})=I_1(0)+(1-\xi)I_2(0).
\end{align}
\begin{align}
I_1(0)&=e^3\int \frac{d^dk}{(2\pi)^d}
\frac{-i\delta _{\mu \nu }}{k^2}(-i\gamma ^{\mu })\frac{i}{k\!\!\!/}(-i\gamma^{\alpha })\frac{i}{k\!\!\!/}(-i\gamma ^{\nu}) \notag \\
&=:-i2e^{3}\gamma ^{\alpha }\frac{1}{(4\pi )^{2}}\frac{1}{\epsilon }.
\end{align}

\begin{align}
I_2(0)&=e^3\int \frac{d^dk}{(2\pi)^d}\frac{-i(-1)(1-\xi)}{k^2}
\frac{k_{\mu}k_{\nu}}{k^2}(-i\gamma^{\mu })\frac{i}{k\!\!\!/} \notag\\
&\times(-i\gamma ^{\alpha})\frac{i}{k\!\!\!/}(-i\gamma ^{\nu})=:
ie^{3}\gamma^{\alpha}\frac{2(1-\xi)}{(4\pi)^2}\frac{1}{\epsilon}.
\end{align}
We then get
\begin{align}
-ie[\Gamma^{A}_{A\psi\psi}]^{\alpha}(0)=:-ie^{3}\gamma^{\alpha}
\frac{2-2(1-\xi)}{(4\pi)^2}\frac{1}{\epsilon }.
\end{align}

\begin{figure}[tbp]
\centering
\scalebox{0.8}{\includegraphics{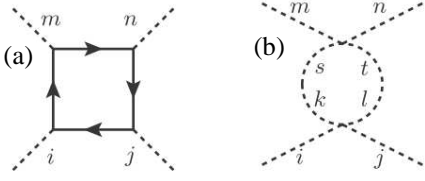}}
\caption{(Color online)The one-loop Feynman diagrams for bosonic self-interaction vertex. The dot line indicates scalar fields propagator, the full line indicates fermion propagator.}
\label{fourv}
\end{figure}
\subsection{Boson self-interaction vertex}
The Feynman diagrams Fig.\ref{fourv} gives the bosonic four-point vertex,
$iV_{ijmn}(p)=iV^{(1)}_{ijmn}(p)+iV^{(2)}_{ijmn}(p)$.
Fig.~\ref{fourv}(a) gives
\begin{align}
&iV^{(1)}_{ij,mn}(0)=-2N_f(-ig)^4\int \frac{d^dk}{(2\pi)^d}
\frac{\text{tr}[\Sigma^i{ik\!\!\!/}\Sigma^j{ik\!\!\!/}\Sigma^m{ik\!\!\!/}\Sigma^n{ik\!\!\!/}]}{k^8}\notag\\
&=:-ig^4\frac{16N_f}{(4\pi)^2}\left[\delta_{ij}\delta
_{mn}+\delta_{im}\delta_{jn}+\delta_{in}\delta_{jm}\right] \frac{1}{\epsilon}.
\end{align}
Fig.~\ref{fourv}(b) gives
\begin{align}
&iV^{(2)}_{ij,mn}(p)=\frac{1}{2}(-i\lambda/3)^{2}(\delta_{ij}\delta_{kl}+\delta_{ik}\delta _{jl}+\delta_{il}\delta_{jk})\notag \\
&\times(\delta_{mn}\delta_{st}+\delta_{ms}\delta_{nt}+\delta_{mt}\delta_{ns})\notag \\
&\times\int \frac{d^dk}{(2\pi)^d}\frac{i\delta_{ks}}{(k+p)^2-m^2}
\frac{i\delta_{lt}}{k^2-m^2},
\end{align}
where we have used the Feynman rule for $O(N_b)$ bosonic self-interaction vertex
\begin{align}
\gamma_{ijmn}=(-i\lambda/3)(\delta_{ij}\delta_{mn}+\delta_{im}\delta_{jn}+\delta _{in}\delta_{jm}).
\end{align}
This expression is logarithmically divergent at large momentum. To eliminate the ultraviolet divergence, setting $p=0$, one get the divergence
\begin{align}
\frac{{i\lambda}^2}{9(4\pi)^2}[(N_b+4)\delta_{ij}\delta_{mn}+2\delta_{im}\delta_{jn}+2\delta_{in}\delta_{jm}]\frac{1}{\epsilon }.
\end{align}
By including all the crossed diagrams, the total correction is given by
\begin{align}
iV^{(2)}_{ijmn}(0)=:\frac{{i(N_b+8)\lambda}^2}{9(4\pi)^2}[\delta_{ij}\delta_{mn}
+\delta_{im}\delta_{jn}+\delta_{in}\delta_{jm}]\frac{1}{\epsilon }.
\end{align}
\section{Renormalization conditions}\label{appendixb}
To determine the counter terms, we expand the renormalization constants as $Z_X=1+\delta_X$, these $\delta_X$ factors are called counter terms. In the perturbation theory, the counter terms are all infinite and they are determined to subtract off the divergence. To this end, we have the following renormalization conditions:
\begin{align}
&-i\Sigma^{\phi}_{\psi\psi}-i\Sigma^{A}_{\psi\psi}+ip\!\!\!/\delta_{\psi}=0,\label{B1} \\
&-i\Sigma^{\psi}_{\phi\phi}-i\Sigma^{\phi}_{\phi\phi}+i[p^2\delta _{\phi}-m^2(\delta_{\phi}+\delta_{m^2})]=0, \\
&i(\delta^{\mu\nu}p^2-p^{\mu}p^{\nu})\Pi(p)-i(\delta^{\mu\nu}p^2-p^{\mu }p^{\nu})\delta_A=0.\label{B3}
\end{align}
\begin{align}
&-ig[\Gamma^{\phi}_{\phi\psi\psi}]^b-ig[\Gamma^{A}_{\phi\psi\psi}]^b-ig\Sigma^{b}\delta_2=0,\label{B4} \\
&-ie[\Gamma^{\phi}_{A\psi\psi}]^{\mu}-ie[\Gamma^{A}_{A\psi\psi}]^{\mu}-ie\gamma^{\mu}\delta_1=0, \\
&iV^{(1)}_{ijmn}(0)+iV^{(2)}_{ijmn}(0)-i(2\delta_{\phi}+\delta_{\lambda})\lambda/3 \notag \\
&\times[\delta_{ij}\delta_{mn}+\delta_{im}\delta_{jn}+\delta_{in}\delta_{jm}]=0 \label{B6}.
\end{align}
Eqs.(\ref{B1})-(\ref{B3}) correspond to fermion self-energy, boson self-energy and gauge self-energy, respectively;
Eqs.(\ref{B4})-(\ref{B6}) correspond to Yukawa vertex, QED$_3$ vertex and boson self-interaction vertex, respectively.
We also define $Z_1=Z_{\psi}Z_e\sqrt{Z_A}$ and $Z_2=Z_{\psi}Z_g\sqrt{Z_{\phi}}$.
These renormalization constants can be read off from these conditions.
\begin{align}
&Z_{\psi }=1-\frac{g^2N_b+2e^2-2(1-\xi )e^{2}}{(4\pi )^{2}\epsilon},\\
&Z_{\phi}=1-\frac{4N_fg^2}{(4\pi)^2\epsilon}.
\end{align}
\begin{align}
&Z_{A}=1-\frac{8N_fe^2}{3(4\pi )^2\epsilon},\\
&Z_{m^2}=1+\frac{(N_b+2)\lambda/3 +4g^2N_f}{(4\pi )^2\epsilon}.
\end{align}
\begin{align}
&Z_1=1-\frac{g^2N_b+2e^2-2(1-\xi)e^2}{(4\pi)^2\epsilon},\\
&Z_2=1+\frac{2(2-N_b)g^2-8e^2+2(1-\xi)e^2}{(4\pi )^2\epsilon}.
\end{align}
\begin{align}
&Z_{\lambda }=1+\frac{8N_fg^2+(N_b+8)\lambda/3-48N_fg^4/\lambda}{(4\pi)^2\epsilon},\\
&Z_e=1+\frac{4N_fe^2}{3(4\pi)^2\epsilon},\\
&Z_g=1+\frac{\left(2N_f-N_b+4\right)g^2-6e^2}{(4\pi)^2\epsilon}.
\end{align}
The beta function, for example $\beta_{\lambda}$, is determined by $d[Z_{\lambda}\mu^{\epsilon}\lambda]/d{\mu}=0$, yielding
\begin{equation}
\beta_{\lambda}=-\epsilon \lambda +\frac{8N_fg^2\lambda }{(4\pi )^2}+\frac{\lambda^2(N_b+8)/3}{(4\pi )^2}-\frac{48N_{f}g^4}{(4\pi )^2}.
\end{equation}
\end{appendix}

\end{document}